\documentclass[prb,superscriptaddress]{revtex4}
\usepackage[utf8]{inputenc}
\usepackage{bm}
\usepackage{graphicx}
\usepackage{amsmath}
\newcommand{\ket}[1]{\left|#1\right\rangle}
\newcommand{\avLambda}[1]{\overline{\Lambda_{#1}}}

\begin{document}
\author{Kirill \surname{Dubovitskii}}
\affiliation{Moscow Institute of Physics and Technology, Dolgoprudny, Russia}

\author{Yuriy \surname{Makhlin}}
\affiliation{Condensed-matter physics Laboratory, HSE University, Moscow, Russia} 
\affiliation{Landau Institute for Theoretical Physics, Chernogolovka, Russia}

\title{Partial randomized benchmarking}

\begin{abstract}
In randomized benchmarking of quantum logical gates, partial twirling can be used for simpler implementation, better scaling, and higher accuracy and reliability.
For instance, for two-qubit gates, single-qubit twirling is easier to realize than full averaging.
We analyze such simplified, partial twirling and demonstrate that, unlike for the standard randomized benchmarking, the measured decay of fidelity is a linear combination of exponentials with different decay rates (3 for two qubits and single-bit twirling).
The evolution with the sequence length is governed by an iteration matrix, whose spectrum gives the decay rates.
For generic two-qubit gates one slowest exponential dominates and characterizes gate errors in three channels. Its decay rate is close, but different from that in the standard randomized benchmarking, and we find the leading correction. Using relations to the local invariants of two-qubit gates we identify all exceptional gates with several slow exponentials and analyze possibilities to extract their decay rates from the measured curves.
\end{abstract}

\maketitle
	

\section{Introduction}

Efficient quantum computers will require calibration of quantum gates to evaluate the effects of environment and noise and to enable quantum-error correction~\cite{QECShor,QECSteane,TerhalRMP15}. Because of the requirements of the threshold theorem for fault-tolerant quantum computations~\cite{FTShor,FTAharonov,FTKitaev,FTKnill}, it is important to quantitatively describe such errors.
Various approaches have been suggested in order to characterize the gates. In general, a non-ideal quantum gate is a superoperator on density matrices rather than a unitary, and its deviation from the ideal gate is described by a collection of numbers, which grows very fast with the number of qubits.
The direct quantum process tomography~\cite{ChuangNielsen97,PoyatosCiracZoller97,MokhseniLidar08} gives a full description.
While the gate set tomography~\cite{NielsenGST21,Petta22,VdspGST22,Siddiqi21} can already be performed on two qubits, for $n$ qubits a complete description of a noise model without any assumptions about its structure requires $O(2^{4n})$ parameters, so that it quickly becomes intractable~\cite{EmersonLaflamme07,Eisert18} in spite of further improvements like the compressed sensing~\cite{FlammiaEisert12,GrossEisertPRL10}.

Randomized benchmarking (RB)~\cite{Emerson2005, Knill2008, Magesan2011} has a more advantageous scaling
and provides a single overall metric for the error level instead of full characterization of its structure.
Random gate sequences are tested, and the sequence fidelity decays exponentially with its length. This allows one to enhance and measure small gate errors in current and prospective quantum-information devices and also makes it robust to state preparation and measurement errors~\cite{MagesanGambettaEmersonPRA12}.

While RB characterizes a set of unitary gates, a specific gate can be tested using the {\it interleaved} RB~\cite{IRBSteffen2012,Gaebler07,SteffenPRA13} (IRB), where a random gate sequence is interleaved with applications of this specific gate. Effectively, this implies averaging of the tested gate over unitary rotations, the so called twirl operation. The resulting averaged gate is a depolarizing channel~\cite{Emerson2005}, characterized by a single number, which can be measured to quantify the fidelity of the tested gate.
Instead of impractical sampling of random gates from the full unitary group, standard RB procedures rely on random Clifford gates, since the Clifford twirl can be substituted for a full unitary twirl~\cite{DVDTerhal02}, because the Clifford group $C_n$ is a 2-design~\cite{DVDTerhal02,Dankert06,MagesanGambettaEmersonPRA12}.

With the current advance of experimental techniques, the Clifford randomized benchmarking can be routinely performed on a small number of qubits~\cite{HomeWineland2009,RiebeBlatt2006}, and the question of interest is extension of these techniques to larger qubit systems. This is not straightforward since the complexity grows, albeit polynomially, with the number of qubits (even though one only needs to sample from the fast growing Clifford group),
because of the gate compilation complexity~\cite{MagesanGambettaEmersonPRA12,Flammia14}. The Clifford RB is widely used for one-two qubit systems, but experiments even with three qubits are rare~\cite{3qubitRBGambetta}, see discussion in Ref.~\cite{Proctor18}.

Various generalizations and modifications of RB are discussed in the literature as well as research to better understand the accuracy of fidelity estimates from RB under various conditions.
Apart from developments of IRB, for example, direct RB with application of many gates in parallel was studied~\cite{Proctor18} and the cross-entropy benchmarking (XEB) was demonstrated~\cite{GoogleSupremacy}.
Here we analyze twirling over a subgroup of the full Clifford (or unitary) group in order to address certain problems of the standard IRB and to simplify and optimize the RB protocol~\cite{BrownEastin}. Specifically, we consider $C_1^{\otimes n}$-twirls, i.e., the use of random single-qubit gates only~\cite{EmersonLaflamme07}. This problem is of interest for current and future analysis of quantum benchmarking.
On one hand, in this approach the gate-compilation problem does not arise, facilitating scaling to larger systems.
Furthermore, already for two qubits the use of random single-qubit instead of two-qubit gates ($C_1^{\otimes 2}$-twirl discussed below) saves resources since the former are typically faster and have higher fidelity, which improves the accuracy (confidence interval) of gate-fidelity estimates.
In contrast, comparison against non-interleaved sequences in $C_2$ (or $C_n$) IRB is only approximate, since contributions of errors in random Cliffords and the interleaved gate do not necessarily add up unless one of these error channels is depolarizing~\cite{IRBSteffen2012,Epstein14}. This increases the confidence interval of the fidelity estimate in IRB~\cite{IRBSteffen2012,Epstein14}.

These remarks are of special importance for experiments with logical qubits (built of several physical qubits in quantum-error correction architectures), since they may be challenging, and high-quality two-qubit gates may be in short supply, so that any overhead here may be a concern  for some time~\cite{BrownEastin}.
Further, $C_1^{\otimes2}$-twirling may be convenient if one tests an idle or a single-qubit gate on two qubits in order to directly address decoherence or cross-talk and spurious interactions~\cite{SteffenPRL12}. Moreover, the standard $C_2$ IRB was found to perform poorly in this kind of setting, when the interleaved gate is of higher fidelity than the random Cliffords~\cite{Epstein14}.

While most experiments implement the full IRB~\cite{Gaebler07,SteffenPRA13}, $C_1^{\otimes 2}$-twirling was also used in some cases~\cite{Martinis,Marcus}.
$C_1^{\otimes n}$-twirls would simplify the procedure, take less resources, and allow to use RB easier even on pairs of qubits including within larger systems and as a part of more complex manipulations. 
The need for less resource-intensive nature of such partial twirling becomes more acute for multi-qubit systems, and the use of our analysis of partial twirls here may be promising, although it requires further developments, in particular, of the theory of local invariants to complete exhaustive analysis of regular and exceptional gates (with one or more dominating decay factors, see below). Nevertheless, it is important to realize that such partial twirling is not universally effective for all logical gates and be able to understand when it is useful and when not.

It is apparent from the experimental data~\cite{Martinis,Marcus} and expectations that this approach based on the single-qubit subgroup produces substantial twirling. However, it is not necessarily complete, and the question arises, which information it provides.
We analyze such {\it partial} twirls for the case of two qubits in this article. We note that the effect of the $C_1^{\otimes n}$-twirl, with averaging over single-qubit unitaries, was studied in Ref.~\cite{EmersonLaflamme07} and for the case of simultaneous RB in Refs.~\cite{SteffenPRA13,SteffenPRL12}.
Related questions for multiqubit systems were addressed recently with a discussion of RB generalizations~\cite{Wehner,Erhard,3qubitRBGambetta}.
For Clifford-subgroup twirling, it was shown~\cite{BrownEastin} that in general the space of qubit density matrices is decomposed into independent blocks with different decay factors in these blocks (cf.~the discussion of three decay factors~\cite{SteffenPRL12}).

The approach, developed in the current article, allowed us not only to analyze the benchmarking of generic gates but also to find all exceptional gates and fully analyze the IRB in all these exceptional cases with the possibility to extract more information from a simpler experimental procedure.
We demonstrate that the dynamics of the noise-averaged evolution operator as a function of the length of the RB gate sequence can be described with a linear markovian operator. We find this operator explicitly, using local invariants of the gate. Its eigenvalues provide the decay rates for the measured fidelity in partial-RB experiments, and accordingly, the generic IRB decay curve is a linear combination of several exponentials, cf.~\cite{BrownEastin,SteffenPRA13,Wehner}.
For generic two-qubit gates, the fidelity decay is dominated by a single exponential, which is close, but not identical, to the result of the standard RB, and we find the deviation of these quantities. Furthermore, we complete the exhaustive analysis by finding all exceptional gates, when more than one exponent is visible in the decay curves, and demonstrate how the RB fidelities can be extracted from the data in these cases. In the opposite limit, we find a family of two-qubit gates, which can be viewed as especially suited for partial RB, since only one decay factor is non-vanishing.
The presented analysis can be generalized to situations with more qubits or larger subsystems ($C_1\otimes C_2$, $C_1^{\otimes 3}$ twirls etc.).

\section{Full interleaved randomized benchmarking}

Here we briefly summarize some basic properties of the RB procedure, needed for our analysis of the partial IRB below.
Detailed accounts can be found, e.g., in Refs.~\cite{Emerson2005,MagesanGambettaEmersonPRA12,Dankert06,MeierThesisColorado2013}.
In particular, we introduce some notation and describe certain assumptions, some customary for RB and some used in this paper to enhance the focus on basic properties of the partial RB.
Consider an arbitrary gate $W$ to be tested. For an ideal unitary operation $w_0$ it acts on density matrices by conjugation, $W_0[\rho]=w_0\rho w_0^\dagger$. We assume that ideally, $W$ should realize a (unitary) gate $W_0$, but due to errors $W=W_0\Lambda$ with the {\it error} superoperator $\Lambda$, close to $\hat1$ for weak errors.
We imply that $W=W_0\Lambda$ is the average over realizations of noise (sometimes $W$ refers instead to a specific realization of the gate which should be clear from the context).

Interleaved randomized benchmarking (IRB) studies sequences of the type
\begin{equation}
\label{seq}
\Lambda_n = FWV_n\dots WV_3 WV_2 WV_1\,,
\end{equation}
with random gates $V_i$, which are sampled uniformly from the relevant group, for instance, in the Haar measure from the unitary group U($d$) or, equivalently~\cite{DVDTerhal02} (i.e. with the same average $\Lambda_n$), from the Clifford group. This equivalence relies on the Clifford group being a 2-design, see also~\cite{MeierThesisColorado2013,3-design,MagesanGambettaEmersonPRA12}. For smaller, or other, subgroups of U($d$) the averaging may be only partial.

The final gate $F\equiv (W_0V_n\dots\allowbreak W_0V_3\allowbreak W_0V_2 W_0V_1)^\dagger$ in Eq.~(\ref{seq}) is chosen such that in the absence of errors the product $\Lambda_n$ reduces to $\hat1$, and any deviation from the identity indicates errors. One typically measures the resulting state after the action of the sequence $\Lambda_n$ on an initial state $\rho_0$. Due to errors the probability to find the system in the state $\rho_0$ decays with the sequence length $n$. The decay rate quantifies the fidelity of the gate $W$. This procedure implies multi-fold repetition of the experiment.

There is a number of factors that influence the resulting operator $\Lambda_n$. Noise and inaccuracies during each instance of $W$ in the sequence~(\ref{seq}) force it to deviate from $W_0$. Here we assume that these noise contributions are uncorrelated for different instances of $W$ even within each sequence (short noise correlation time). The total error of the sequence contains independent contributions from all terms in Eq.~(\ref{seq}), and averaging over repetitions results in replacement of each $W$ by its noise-averaged value, $W_0\Lambda$. Moreover, we assume that the random gates $V_i$ are error-free for the purposes of this paper.
In principle, in the standard IRB errors in $V_i$'s are accounted for by subtracting the decay constant for non-interleaved sequences of only $V_i$'s~\cite{IRBSteffen2012} (though accuracy of this approach is proven only for depolarizing noise in $V_i$~\cite{Epstein14}). We neglect this contribution since we focus here on a different phenomenon; this is especially justified for the case of most interest, when $W$ is a two-qubit gate, while $V_i$'s are single-qubit gates, typically, with much lower errors.

Since $W=W_0\Lambda$, one can rewrite (\ref{seq}) as a product of conjugate $\Lambda$'s:
\begin{equation}
\Lambda_n =
F W_0\Lambda V_n \dots W_0\Lambda V_3 W_0\Lambda V_2 W_0\Lambda V_1
=
(U_n^\dagger \Lambda U_n) (U_{n-1}^\dagger \Lambda U_{n-1})
\dots
(U_2^\dagger \Lambda U_2) (U_1^\dagger \Lambda U_1)
\,,
\label{seqU}
\end{equation}
with the unitaries $U_i$, related to $V_i$ via:
\begin{equation}\label{eq:indprod}
V_1=U_1,\quad
V_iW_0 = U_iU_{i-1}^\dagger \textrm{ for }i=2\dots n \,.
\end{equation}
(We assume that we deal with $n$ equivalent implementations of the same gate $W$.)
Uniform distribution for $V_i$'s over the (unitary/Clifford or other) group implies the same for $U_i$'s (since $W_0$ belongs to the same group; this is not the case for partial RB). Hence averaging of $\Lambda_n$ over the random $V_i$'s reduces to independent averaging of each $U^\dagger\Lambda U$-term on the rhs of Eq.~(\ref{seqU}).
Every such term, a superoperator $\Lambda_U$, maps a state $\rho$ to
\begin{equation}\label{eq:LamU}
\Lambda_U[\rho]\equiv U^\dagger \Lambda[U\rho U^\dagger]U
\,.
\end{equation}
Its average over realizations of the random $U$ (we use both notations, the over-bar and the angular brackets) is:
\begin{equation}\label{eq:ave}
\bar \Lambda[\rho] \equiv \left\langle\Lambda\right\rangle [\rho] =
\langle \Lambda_U[\rho]\rangle_U \,.
\end{equation}
Here the last subscript $U$ denotes averaging over the group.
One can see that the averaged gate $\bar \Lambda$ is {\it isotropic} with respect to the group rotations, i.e., invariant under an arbitrary basis change $L$ from the group:
\begin{eqnarray}
\bar \Lambda[L\rho L^\dagger] = L\bar \Lambda[\rho]L^\dagger \,,
\end{eqnarray}
because according to Eq.~(\ref{eq:LamU}) for any $L$ we have $\Lambda_U[L\rho L^\dagger] = L \Lambda_{UL}[\rho] L^\dagger$, which can be immediately averaged over $U$ (or equivalently, over $U'=UL$).

The isotropy strongly constrains the degrees of freedom in the gate $\bar \Lambda$, making it a depolarizing channel:
\begin{eqnarray}
\label{eq:al}
\bar \Lambda[\rho] = (1-\mu) \frac{\hat 1}{d}
+ \mu \rho \,,
\end{eqnarray}
where $d$ is the dimension of the Hilbert space. It interpolates between the identity map at $\mu=1$ and the completely depolarizing channel, a constant map to $\hat 1/d$ at $\mu=0$.

Thus, $\avLambda{n}$ is isotropic for any $n$. RB investigates how $\mu_n\equiv\mu(\avLambda{n})$ depends on $n$.
According to Eq.~(\ref{seqU}) one has $\avLambda{n} = (\bar \Lambda)^n$, and using Eq.~(\ref{eq:al}) one finds that $\mu_n=\mu^n$ decays exponentially with the sequence length $n$. By measuring this exponential decay one can extract $\mu$, the fidelity of the tested gate $W$.

\section{Averaging a two-qubit operation over single-qubit gates}

Let us now consider twirls over a smaller group, the single-qubit group $C_1^{\otimes n}$ (or $U_1^{\otimes n}$).
On one hand, randomization with only single-qubit gates appears to be still sufficiently powerful. However, it is not obvious, if it is complete, that is if $\avLambda{n}$ is depolarizing. There is a number of questions, which we analyze below. Does the fidelity of the sequence decay exponentially? If not, what kind of decay is expected and which information about the gate can be extracted from this decay? If one fits the decay curve with an exponential, how is the extracted exponent related to the RB-fidelity of a complete RB experiment (with random unitary or Clifford two-qubit gates $V_i$ and complete averaging)?

We discuss these questions for two-qubit gates with single-qubit randomization as described above and show that the decay is characterized by three exponentials, and then show how to complete the analysis for the case of two qubits. This approach can be extended to a more general situation of partial averaging over a subgroup.

To begin the analysis, note that in the case of partial averaging the considerations of the previous section fail. More specifically, in Eq.~(\ref{eq:indprod})
$U_i$'s do not belong the group unless $W_0$ is a single-qubit gate itself, and averaging over $U_i$'s in Eq.~(\ref{seqU}) cannot be done straightforwardly. Instead, we rewrite the sequence (\ref{seq}) as follows:
\begin{equation}
\label{eq:seq2qb}
\Lambda_n =
\tilde F_n
(\tilde U_n^\dagger W_0 \Lambda\tilde U_n) \dots
(\tilde U_2^\dagger W_0 \Lambda\tilde U_2)
(\tilde U_1^\dagger W_0 \Lambda\tilde U_1)
\,,
\end{equation}
where
\begin{equation}
V_1=\tilde U_1,\quad
V_i= \tilde U_i \tilde U_{i-1}^\dagger \textrm{ for }
i=2\dots n \,,
\end{equation}
and
\begin{equation}\label{eq:defFntilde}
\tilde F_n =
\left[
(\tilde U_n^\dagger W_0\tilde U_n) \dots
(\tilde U_2^\dagger W_0\tilde U_2)
(\tilde U_1^\dagger W_0\tilde U_1)
\right]^\dagger
\,.
\end{equation}
In this case $\tilde U_i$'s are independent single-qubit random gates. Note that they also enter Eq.(\ref{eq:defFntilde}) for the final gate $\tilde F_n$.
However, if $W_0$ is also a single-qubit gate, Eq.~(\ref{seqU}) can be applied, and this is used below in the following subsection.

\subsection{Testing a trivial two-qubit operation}

Let us begin our analysis from the case of $W_0=\hat 1$. Testing the identity gate may probe the influence of noise or decoherence. Then
\begin{equation}
\avLambda{n} = \bar \Lambda ^n \,.
\end{equation}

$\bar \Lambda$ is isotropic w.r.t. single-qubit rotations, or {\it locally invariant}.
It maps a two-qubit density matrix $\rho = \frac{1}{4} + \frac{1}{2}{\bf s}\bm{\sigma}^{(1)} + \frac{1}{2}{\bf p}\bm{\sigma}^{(2)}
+ \beta_{ij} \sigma_i^{(1)} \sigma_j^{(2)}$
to a matrix of the same form. One can easily see that in terms of $\bm{s,p},\beta$ the most general locally invariant mapping is:
\begin{equation}
\bm{s}\mapsto a\bm{s},\quad
\bm{p}\mapsto b\bm{p},\quad
\beta \mapsto c\beta
\end{equation}
with three independent real factors $a,b,c$, which satisfy $|a|,|b|,|c|\le1$. Hence
\begin{equation}
\avLambda{n}:\quad
\bm{s}\mapsto a^n\bm{s},\quad
\bm{p}\mapsto b^n\bm{p},\quad
\beta \mapsto c^n\beta
\,.
\end{equation}

For an operation with given $a,b,c$, if we average it over the whole SU(4), what value of $\mu$ would we obtain? Apparently, $\mu$ would be a linear combination $xa+yb+zc$. Since  for $a=b=c$ they coincide with $\mu$, one finds that $x+y+z=1$.
Furthermore, averaging, e.g., with the gate CZ (or CNOT) replaces $a,b,c$ with $\frac{2}{3}c+\frac{1}{3}a,
\frac{2}{3}c+\frac{1}{3}b,
\frac{5}{9}c+\frac{2}{9}a+\frac{2}{9}b$, but should keep the same $\mu$. All this allows us to find that $x=y=\frac{1}{5}, z=\frac{3}{5}$, and thus:
\begin{equation}
\mu=\frac{a+b+3c}{5} \,.
\end{equation}
This is the value, which the standard randomized benchmarking (with complete averaging over all Clifford or unitary two-qubit gates $V_i$) would measure.

For instance, the initial state 00 has $s_z=p_z=1/2$ and $\beta_{zz}=1/4$, so that the probability to find the same state after $n$ rounds decays as $(1+a^n+b^n+c^n)/4$. From this value one can extract $a$, $b$, $c$. To simplify extraction, one can apply the operation to various initial states and measure probabilities of various final states. For instance, if the system is prepared in the initial state $\ket{\uparrow\uparrow}$, then by repeating the experiment one can measure the probabilities $P_{\uparrow\uparrow}$, $P_{\uparrow\downarrow}$, $P_{\downarrow\uparrow}$, $P_{\downarrow\downarrow}$ of the four computational-basis states after application of the IRB-sequence. From these one can find the three decaying exponentials directly:
\begin{equation}\label{eq:extrabc}
\begin{pmatrix}1&-1&-1&1\\1&-1&1&-1\\1&1&-1&-1\\1&1&1&1\end{pmatrix}
\begin{pmatrix}
P_{\uparrow\uparrow}\\
P_{\uparrow\downarrow}\\
P_{\downarrow\uparrow}\\
P_{\downarrow\downarrow}
\end{pmatrix}
=
\begin{pmatrix}c^n\\b^n\\a^n\\1\end{pmatrix}
\,.
\end{equation}
This allows one to extract three decay factors, $a$, $b$, $c$, separately (they are all close to 1 in the case of small errors).

Obviously, when two qubits are decoupled and uncorrelated, $a$ characterizes single-qubit errors on the first qubit, $b$ describes errors on the second qubit. If only single-qubit errors are present, $c=ab$. Hence, the difference $c-ab$ describes errors associated with interaction/cross-talk between the qubits, or any other kind of correlated noise experienced by them~\cite{SteffenPRL12,SteffenPRA13}.

\subsection{Testing an arbitrary two-qubit gate}

Twirling over single-qubit gates for a two-qubit system is a particular case of averaging over a subgroup (cf.~\cite{BrownEastin}). Another simple example is twirling only by rotations around one axis (say, the $z$-axis) for a single qubit. In such cases, in contrast to twirling over the whole unitary or Clifford group, the twirled operation $\bar\Lambda$ is not necessarily characterized by a single depolarizing parameter $\mu$, but in general by more parameters (see above). For the $z$-twirling of a single qubit, as it happens, we also have three decay factors: in the language of the Bloch sphere,
one for the $z$-component, and two conjugate factors for the $xy$-plane, which results in oscillatory in-plane decay.

As we discussed in the introduction, such partial RB is of special interest, and we analyze which information does one learn from such measurements. While some properties are more general, below we focus on the single-qubit subgroup (either unitary $U_1^{\otimes 2}$ or Clifford $C_1^{\otimes 2}$ with equivalent twirling properties~\cite{DVDTerhal02}).

Let us derive an expression for the averaged operation after many repetitions in the case of the $C_1^{\otimes 2}$ twirling. By rewriting expression (\ref{eq:seq2qb}) for the interleaved sequence, one finds the following recurrence relation for the operation $\avLambda{n}$, the result of the $n$-step interleaved RB (see Appendix~\ref{sec:appRecRel}):
\begin{equation}
\avLambda{n+1} = \left\langle
(W_0^\dagger\avLambda{n} W_0) \cdot \Lambda
\right\rangle
\label{eq:LamRec}
\,,
\end{equation}
where the angular brackets denote averaging over the subgroup (\ref{eq:ave}), while the gate error superoperator $\Lambda$ was defined before Eq.~(\ref{seq}) and coincides with $\Lambda_1$, cf.~also a discussion after Eq.~(\ref{seq}).
Clearly, $\avLambda{n}$ is subgroup-invariant and characterized by the corresponding parameters (three numbers $a_n,b_n,c_n$ both for single-qubit averaging and two-qubit $W$ as well as for $z$-averaging and a single-qubit $W$). This recurrence relation is a central result in the analysis. Our further goal is to solve this recurrence relation.

First, we note that Eq.~(\ref{eq:LamRec}) simplifies when the gate $W_0$ belongs to the group, which we are averaging over. This includes the case of the whole unitary U(2) (or Clifford $C_2$) group (standard RB), and the case of the idle gate $W_0=\hat 1$. In these cases $W_0$ drops out of Eq.~(\ref{eq:LamRec}), and since $\avLambda{n}$ is locally invariant, it can be taken out of the averaging, which implies that $\avLambda{n} = \overline{\Lambda}^n$. However, in general, for an arbitrary groups and gates $W_0$, the average product in (\ref{eq:LamRec}) does not factorize. Below we analyze this expression in this generic situation, for the $C_1^{\otimes2}$ twirl and an arbitrary gate $W_0$.

Each averaged error operator $\avLambda{n}$ is characterized by a triple of numbers, which can be combined into a vector
\begin{equation}
{\bf f}_n \equiv \begin{pmatrix}a_n\\b_n\\c_n\end{pmatrix} \,,
\end{equation}
and Eq.~(\ref{eq:LamRec}) is a linear relation between ${\bf f}_{n+1}$ and ${\bf f}_n$, which we describe by an {\it iteration matrix} $M$:
\begin{equation}
{\bf f}_{n+1} = \hat M {\bf f}_n \,.
\end{equation}
Clearly, the $n$-dependence of ${\bf f}_n$, and hence results of any measurement in an RB experiment, are determined by the eigenvalues of the $3\times3$ matrix $\hat M$. In the following we analyze the spectrum of this matrix.

The error-free (identity) operation $\Lambda=\hat1$ corresponds to ${\bf f}=(1,1,1)$. Thus, neglecting preparation errors, we find that ${\bf f}_n = \hat M^n (1,1,1)$.

While the exact spectrum of $M$ depends on the properties of the error-operator $\Lambda$, in the zero-order approximation, dropping the factor $\Lambda$ on the rhs of Eq.~(\ref{eq:LamRec}), we obtain an error-free iteration matrix $M^0$:
\begin{equation}
M^0:\quad
\avLambda{n+1} = \left\langle
W_0^\dagger\avLambda{n} W_0
\right\rangle
\label{eq:LamRec0}
\,.
\end{equation}
We first find the spectrum of $M^0$, and the spectrum of $M$ can then be found perturbatively in small errors. In particular, the spectra of $M$ and $M^0$ are close (assuming gate errors are weak).

Introducing matrix elements of $W_0$ and $\Lambda$, we have found from Eq.~(\ref{eq:LamRec0}) the matrix elements of $M$ by direct calculation in terms of the matrix elements of $\Lambda$. In particular, in the error-free case $\Lambda=1$, we found certain relations between these elements.
The first set of three relations,
\begin{equation}\label{eq:rel1}
M^0_{i1} + M^0_{i2} + M^0_{i3} = 1 \,,\quad i=1,2,3 \,,
\end{equation}
follows from the fact that an error-free operator $\avLambda{n}$ remains error-free after application of (\ref{eq:LamRec0}). In other words, $(1,1,1)$ is the eigenvector of $M^0$ with eigenvalue 1.

Further, due to trace conservation
\begin{eqnarray}
M^0_{11} + M^0_{21} + 3 M^0_{31} &=& 1 \,,\nonumber\\
M^0_{12} + M^0_{22} + 3 M^0_{32} &=& 1 \,,\label{eq:relTr}\\
M^0_{13} + M^0_{23} + 3 M^0_{33} &=& 3 \,.\nonumber
\end{eqnarray}
Trace conservation here implies that the traces of $\avLambda{n+1}$ and $\avLambda{n}$ coincide, which follows directly from (\ref{eq:LamRec0}).

Relations (\ref{eq:rel1}), (\ref{eq:relTr}) strongly constrain the structure of the iteration matrix $M^0$. However, there is a further relation: the matrix $M^0$ remains intact under qubit transposition as we show in the next section
with the use of local invariants. This implies that $M^0$ is symmetric: $M^0_{ij}$ remains the same if in the subscript $ij$ each 1 is replaced by 2 and 2 by 1, or explicitly
\begin{eqnarray}\label{eq:relSym}
M^0_{12} = M^0_{21} \,,
M^0_{13} = M^0_{23} \,,
M^0_{31} = M^0_{32} \,,
M^0_{11} = M^0_{22} \,.
\end{eqnarray}

Using this and the previous relations, we find the general form of the iteration matrix
\begin{equation}\label{eq:IterMat}
M^0 = \begin{pmatrix}
m_1 & m_2 & 1-m_1-m_2\\
m_2 & m_1 & 1-m_1-m_2\\
\frac{1-m_1-m_2}{3} & \frac{1-m_1-m_2}{3} & \frac{1+2m_1+2m_2}{3} &
\end{pmatrix}
\,.
\end{equation}
and its spectrum:
\begin{eqnarray}\label{eq:Sp}
\textrm{Spectrum of $M^0$: } 1,m_1-m_2,\frac{5m_1+5m_2-2}{3} \,.
\end{eqnarray}
The corresponding eigenvectors (isotropic superoperators) are $W$-independent: $(1,1,1)$ for the identity superoperator, $(1,-1,0)$ and $(3,3,-2)$ for the antisymmetric/symmetric traceless superoperators. Hence iteration matrices $M^0$ for all gates commute.

\begin{figure}
\includegraphics[width=0.5\columnwidth]{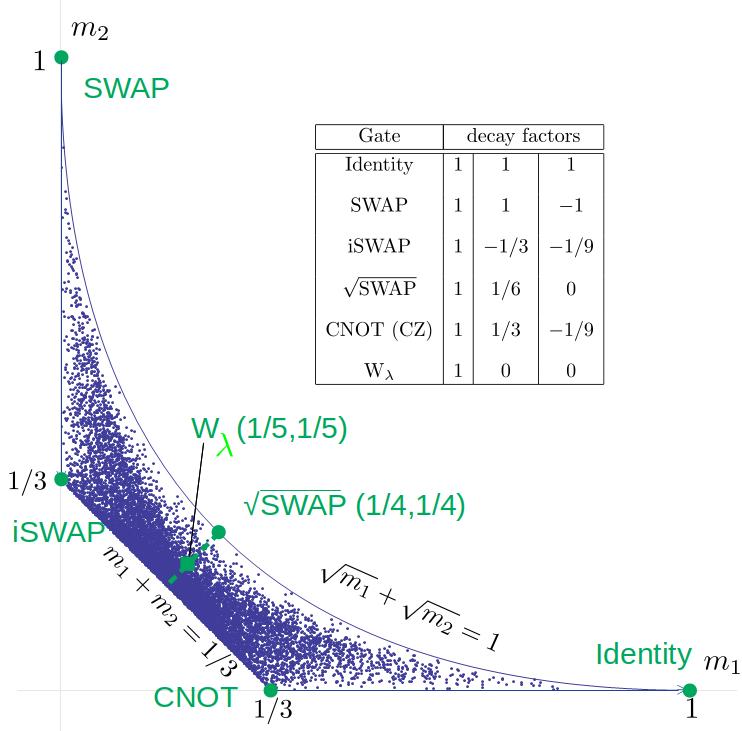}
\caption{Distribution of possible values of the entries $m_1$, $m_2$ of the iteration matrix $M^0$ for possible two-qubit gates, drawn uniformly from the unitary group U(4). Each point $(m_1,m_2)$ corresponds to a family of two-qubit gates. Large solid dots correspond to (the families of) the gates SWAP, iSWAP, $\sqrt{\textrm{SWAP}}$, identity, and CNOT. Boundaries of this region are discussed in the text.
The dashed line indicates the family in Eq.~(\ref{eq:5fam}) with the solid square showing the gate $W_\lambda$ especially suited for partial RB --- with only a single decay factor (here $\cos\lambda\pi=-1/5$), see discussion around Eq.~(\ref{eq:5fam}).	
Inset: the partial-RB decay factors for each of these gates, that is the eigenvalues of the respective iteration matrix $M^0$.}
\label{fig:m1m2}
\end{figure}

There are no further relations between the matrix elements of the iteration matrix as illustrated in Fig.~\ref{fig:m1m2}, in which $(m_1,m_2)$ for various possible gates are plotted. They all fall within an area, limited by the four curves:
\begin{eqnarray}
0\le m_1\le1\,,\quad 0\le m_2\le1\,, \label{eq:cond1m1m2}\\
m_1+m_2 \ge \frac{1}{3}\,,\label{eq:cond2m1m2}\\
\sqrt{m_1}+\sqrt{m_2}\le 1 \,.\label{eq:cond3m1m2}
\end{eqnarray}
These relations were directly checked analytically, for instance, by using an explicit representation of two-qubit gates $W_0$ from all possible equivalence classes with the same local invariants~\cite{Vala2003,Vala2013,Invariants}. In terms of these invariants $G_1$, $G_2$ (cf. the next section) they read:
\begin{eqnarray}
2|G_1|+1\ge|G_2| \,,\\
|G_1|\ge0\,,\\
(G_2)^2+3 \ge 12|G_1| \,.
\end{eqnarray}

\section{Iteration matrix and local invariants}

The special form of Eq.~(\ref{eq:LamRec0}) entails an important property of the mapping and the corresponding matrix $M^0$. Indeed, the mapping from $\avLambda{n}$ to $\avLambda{n+1}$ is the same for two two-qubit gates $w^A_0$ and $w^B_0$ if they differ only by application of single-qubit gates before and/or after the gate: $w^A_0 = S_1S_2w^B_0T_1T_2$, where $S_1$, $S_2$, $T_1$, $T_2$ are single-qubit unitary gates applied to the qubit 1 or 2 as indicated by the subscript. Such two gates are referred to as locally equivalent, or equivalent up to local transformations, and physically have the same correlation, or entanglement properties. It has been shown that two gates are locally equivalent if and only if they have the same value of the so called local invariants, and a complete set of such invariants was found~\cite{Invariants}. These invariants of a two-qubit gate, a complex number $G_1$ and a real number $G_2$, are given by explicit expressions in terms of the matrix of the gate and thus can be efficiently calculated. They turn out to be useful in the analysis of optimal decompositions of complex unitary operations in terms of elementary quantum logic gates for specific physical realizations of qubits~\cite{Plourde2004,Wineland2010,DasSarmagateFidelity}.

Explicitly the invariants can be found using the $4\times4$ matrix $Q$ of transformation to the Bell basis~\cite{Invariants}: first, one finds the matrix of the gate in the Bell basis, $w_B=Q^\dagger wQ$, then the product $\omega=w_B^Tw_B$, and finally, for a unitary gate the invariants are given by
\begin{equation}
G_1 = \frac{{\mathop{\textrm{tr}}}^2\omega}{16},\quad
G_2 =
\frac{{\mathop{\textrm{tr}}}^2\omega -\mathop{\textrm{tr}}\omega^2}{4}
\,.
\end{equation}

The observation above about the local invariance of the iteration matrix $M^0$ implies that it is completely determined by the local invariants $G_1$, $G_2$. This observation has useful consequences for our analysis. First, one can show explicitly that the gates $W_0$ and SWAP$\cdot W_0\cdot$SWAP, which differ only by the transposition of the two qubits, have the same values of the local invariants. Hence they are locally equivalent and, in particular, have the same iteration matrix $M^0$. Here the SWAP gate is a standard operator, which exchanges the states of two qubits. On the other hand, it is obvious, and can be checked directly, that the $M^0$-matrices for $W_0$ and SWAP$\cdot W_0\cdot$SWAP differ by the exchange of the first two basis vectors, that is by the transposition of the first two columns and first two rows. This immediately proves the relation (\ref{eq:relSym}).

Further, we found an explicit expression for the iteration matrix $M^0$ (\ref{eq:IterMat}) in terms of the local invariants:
\begin{eqnarray}\label{eq:mviaG}
m_1 &=& \frac{2|G_1|+G_2+1}{6} \,,\\
m_2 &=& \frac{2|G_1|-G_2+1}{6} \,.
\end{eqnarray}
One can verify these identities by various methods. For example, we used the fact that the family of the gates $W_0=\exp(\frac{i}{2}[c_x\sigma_x^1\sigma_x^2 + c_y\sigma_y^1\sigma_y^2 + c_z\sigma_z^1\sigma_z^2])$, with real $c_{x,y,z}$ contains representatives with all possible local invariants (hence, any two-qubit gate is locally equivalent to some gate in the family)~\cite{Vala2003}. For this family we calculated both sides of these identities in order to verify them.

Thus, we found that the iteration matrix $M^0$ is completely determined by the local invariants of the gate $W_0$. However, we notice that the matrix depends only on $G_2$ and the absolute value $|G_1|$, but not on the phase of $G_1$. Thus, a question arises which two-qubit gates form this fixed-iteration-matrix family with the same $|G_1|$ and $G_2$, but various $\mathop{\textrm{arg}}G_1$.
To analyze it, it is convenient to introduce the parameters $k_x=\cos 2c_x$, $k_y=\cos 2c_y$, $k_z=\cos 2c_z$. In the $k$-space each point in the cube $-1\le k_x,k_y,k_z\le1$ represents two local-equivalence classes of two-qubit gates with the same $G_2$ and complex conjugate values of $G_1$.
One can verify that
\begin{equation}
G_2=k_x+k_y+k_z \quad\textrm{ and }\quad |G_1|=(2+G_2^2-{\bf k}^2)/8 \,.
\end{equation}
(In fact, $\mathop{\textrm{Re}}G_1 = (k_x+k_y+k_z+k_xk_yk_z)/4$ and $\mathop{\textrm{Im}}G_1 = \pm\sqrt{(1-k_x^2)(1-k_y^2)(1-k_z^2)}/4$.)
Thus, the family with fixed $|G_1|$ and $G_2$ is a circle in $k$-space, orthogonal to the main diagonal $(1,1,1)$ and with a center on this diagonal.

More precisely, the family only covers the part of circle within the cube. At the same time, $\mathop{\textrm{arg}}G_1$ does not assume all values in $[0,2\pi]$ on this circle.
Using the expression~\cite{Vala2013} for the uniform (Haar) measure in terms of the local invariants $G_1$, $G_2$, one finds that it can be expressed as $\propto dm_1 dm_2 d(\mathop{\textrm{arg}}G_1)$, and thus the density of points in Fig.~\ref{fig:m1m2} shows directly, which fraction of the full interval $[0,2\pi]$ is covered by admissible values of $\mathop{\textrm{arg}}G_1$ for given $|G_1|$ and $G_2$. In particular, at each point on the main diagonal of the cube in $k$-space the phase $\mathop{\textrm{arg}}G_1$ has only one fixed value, and thus the density of points at the upper boundary in Fig.~\ref{fig:m1m2} vanishes.
We note further that the circles for the identity and the SWAP gates are just the points $(1,1,1)$ and $(-1,-1,-1)$ on the main diagonal.

\section{Decay factors}

The spectrum of $M$ determines the decay factors in the decay curves measured in an IRB experiment as we discussed above. It is close to the spectrum of $M^0$, and as one can see from the explicit expression (\ref{eq:Sp}), one eigenvalue of $M$ is always close to 1, while two other eigenvalues are typically smaller (it follows from Eq.~(\ref{eq:Sp}), Eqs.~(\ref{eq:cond1m1m2})-(\ref{eq:cond3m1m2}) and Fig.~\ref{fig:m1m2} that the eigenvalues cannot exceed 1 by absolute value; they always correspond to decay rather than growth.) This implies that the measured decay curve is a linear combination of three decaying exponentials, one slow and two others, which decay fast and vanish already at small values of the sequence length $n$. Below in this section we compare this slow decay constant to that observed in standard IRB with complete twirling over SU(4) or $C_2$, see Eq.~(\ref{eq:mu2order}) and below.

However, we begin with the analysis of exceptions to this general picture: our approach allows us to find and analyze all exceptional cases, when more than one exponent with the decay constant close to $\pm1$ may appear so that more than one exponential is visible in the decay curve. Analysis, based on the explicit expressions~(\ref{eq:Sp}) for the spectrum, demonstrates that this happens only near $m_1=1,m_2=0$ and $m_1=0,m_2=1$. Translation to the language of local invariants allows us to find that the exceptional gates are those close to the identity and the SWAP gate, as well as to those locally equivalent to them. In other words, the exceptional cases are single-qubit gates, perhaps, in combination with one SWAP gate.

Thus, generically only one decay factor defines the IRB decay curve. As for the exceptional situations, for nearly single-qubit gates there are three decay factors $a,b,c$, close to 1, which determine the decay, and they can be extracted, for instance, using the procedure described near Eq.~(\ref{eq:extrabc}). In the remaining exceptional case of the SWAP and locally equivalent gates,
the eigenvalues of $M$ are close to those of $M^0$ $(1,1,-1)$ and can be extracted similarly as above. In this case the iteration matrix is
\begin{equation}
M = \begin{pmatrix}0&a&0\\b&0&0\\0&0&c\end{pmatrix}
\,,
\end{equation}
where $a$, $b$, $c$ can be expressed via matrix elements of the $\Lambda$ superoperator in the basis of Pauli matrices $\hat\sigma_i\otimes\hat\sigma_j$ in the space of density matrices: if we define the basis vectors as $\sigma_i\otimes\sigma_0$, $\sigma_0\otimes\sigma_j$, $\sigma_i\otimes\sigma_j$, $\sigma_0\otimes\sigma_0$ (with $i,j=x,y,z$), then
\begin{equation}
a = \frac{1}{3} \sum_{\alpha=1,2,3} \Lambda_{\alpha\alpha}\,,
b = \frac{1}{3} \sum_{\alpha=4,5,6} \Lambda_{\alpha\alpha}\,,
c = \frac{1}{9} \sum_{\alpha=7}^{15} \Lambda_{\alpha\alpha}\,.
\end{equation}

Then we find that after $n$ steps the error superoperator $\avLambda{n}$ is described by ${\bf f}_{n} = ( (ab)^{n/2}, (ab)^{n/2}, c^n)$ for even $n$ and
${\bf f}_{n} = (a(ab)^{(n-1)/2}, b(ab)^{(n-1)/2}, c^n)$ for odd $n$.
This gives us a simple procedure for extracting all three decay factors, $a$, $b$, $c$ from combinations in Eq.~(\ref{eq:extrabc}):
\begin{equation}\label{eq:extrabcSWAP}
\begin{pmatrix}1&-1&-1&1\\1&-1&1&-1\\1&1&-1&-1\\1&1&1&1\end{pmatrix}
\begin{pmatrix}
P_{\uparrow\uparrow}\\
P_{\uparrow\downarrow}\\
P_{\downarrow\uparrow}\\
P_{\downarrow\downarrow}
\end{pmatrix}
=
\begin{pmatrix}c^n\\a^kb^{k+p}\\a^{k+p}b^k\\1\end{pmatrix}
\quad\textrm{ for }n=2k+p,\ p=0/1
\,.
\end{equation}
Here the powers on the rhs depend on the parity $p=0/1$ of the length sequence $n$.
Thus, one can easily extract the decay factors of the gate by using the fact that the upper entry of the column (\ref{eq:extrabcSWAP}) is multiplied by $c$ with extension of the random sequence by one step, while the second and third entries are multiplied alternatively by $a$ and $b$ on odd and even steps. Having extracted $a$, $b$, and $c$, one can also obtain the decay factor $\mu=(a+b+3c)/5$, which would be measured in the standard, more complex IRB experiment with complete twirling.

Now let us account, perturbatively, for the deviations of the error operator $\Lambda$ from identity: $\Lambda_{\alpha\beta}=\delta_{\alpha\beta} + \epsilon_{\alpha\beta}$ with $\epsilon_{\alpha\beta}\ll1$.
We find for the highest eigenvalue of $M$ in the non-degenerate case:
\begin{equation}\label{eq:mu2order}
\mu = 1 + \frac{1}{15}\mathop{\textrm{Tr}}\hat \epsilon =
\frac{1}{15}\mathop{\textrm{Tr}}\hat \Lambda \,.
\end{equation}
We note that exactly this value (to the first order in $\epsilon$) one would obtain in a full RB procedure with complete averaging over all random two-qubit operations.
Thus, the highest eigenvalue of $M$ coincides with the decay factor for the full RB, to the first order in the errors $\epsilon$. This means that the decay factor extracted from the partial RB experiment would coincide with that in the full RB, and hence the simpler partial twirling is efficient.

To study this further, we  found the second-order correction to the highest eigenvalue $\mu$ in the case of the error operator $\Lambda$, isotropic w.r.t. single-qubit rotations and described by the parameters $a$, $b$, $c$ close to 1:
\begin{equation}\label{eq:2ndorder}
\mu = \frac{a+b+3c}{5}
+ \frac{(m_1-m_2)(a-b)^2}{10(1-m_1+m_2)}
- \frac{3(-2 +5m_1+5m_2)(a+b-2c)^2}{250(-1+m_1+m_2)}
\,.
\end{equation}
Thus, due to these corrections a difference appears between the results of the full RB and simple partial RB (with single-qubit twirling only). Hence strictly speaking the partial RB does not reproduce the result of full RB, although corrections are typically weak.
Note that the second-order correction diverges at the exceptional gates (locally equivalent to the identity or SWAP); thus, for gates $W$ close to such exceptional gates deviations between full and partial RB are stronger.

We note also a special gate family with $m_1=m_2=1/5$ especially suited for partial randomized benchmarking with only one non-vanishing decay constant, see Appendix~\ref{sec:appGateFamily}.

\section{Conclusions}

We analyzed the process of partial randomized benchmarking with the focus on the case of testing a two-qubit quantum gate with twirling only over single-qubit rotations (in other words, with interleaving only with random single-qubit gates). We demonstrated that in this case, unlike for the standard randomized benchmarking, the decay of the fidelity as a function of the length of the gate sequence is not purely exponential, but is a combination of three exponential contributions with three different decay factors. These three exponents can be extracted from the experiment and provide information about the errors of the tested quantum gate.

To analyze these decay factors, we showed that the dynamics of the realization-averaged RB sequence as a function of its length may be described as linear and markovian with the use of a $3\times3$ iteration matrix $M$.

In the absence of errors, we found a complete description of the iteration matrix $M^0$. We expressed it in terms of the local invariants of the tested two-qubit gate. This allows one to efficiently find the matrix and its spectrum for a given gate, and thus to analyze partial-RB experiments.

It turns out that for generic gates only one of the three decay factors is close to 1 in absolute value, while the other two are smaller. As a consequence, already for not too long sequences only one exponential survives, and the experimental dependence of fidelity on the sequence length is just exponential to a high accuracy (we even found a family~(\ref{eq:5fam}) of two-qubit gates, especially suited for partial RB, where the second and third decay factors vanish). Furthermore, the decay factor of this exponential is very close to that, which one would obtain in a full RB experiment with complete twirling. Thus, a simplified partial RB provides the same information as the standard full-scale RB, which is harder to implement experimentally.

However, we found out that there are corrections to this statement: while the slowest decay factor in partial RB coincides with the full-RB decay factor to the leading order in the size of the errors, the second-order corrections are non-zero, and we found explicit expressions.

Moreover, there exist exceptional quantum gates: for these gates, more than one of the three decay factors are close to one in absolute value, and hence the decay curve is not a simple exponential. Using the local invariants of the gates, we found and analyzed all the exceptional gates. These gates are the identity, SWAP, and all the gates, locally equivalent to these two. The gates close to these also have similar properties.

Three decay factors for a generic gate, if measured, can be viewed as a fingerprint of the tested two-qubit gate. This fingerprint determines the gate, as we showed, up to local single-qubit operations and up to the phase of the invariant $G_1$. It defines not a unique gate but a one-parameter family of gates (or rather, a family of local equivalence classes).

\appendix

\section{Recurrence relation for the error superopertor}
\label{sec:appRecRel}

Here we derive the recurrence relation (\ref{eq:LamRec}). We start from Eq.~(\ref{eq:seq2qb}) and use it to relate $\Lambda_{n+1}$ to $\Lambda_n$:
\begin{equation}\label{eq:apRec}
\Lambda_{n+1} (\tilde U_1,\ldots, \tilde U_{n+1})
= (\tilde U_1^\dagger W_0^\dagger \tilde U_1)
\Lambda_n(\tilde U_2,\ldots, \tilde U_{n+1})
(\tilde U_1^\dagger W_0 \Lambda \tilde U_1)
\,.
\end{equation}
To find $\avLambda{n+1}$ from this, we need to (i) average over
$\tilde U_2$, \dots, $\tilde U_{n+1}$ (that is average each of them over the group) and (ii) average then over $\tilde U_1$ (averaging can of course be done in any order). Averaging in stage (i) transforms $\Lambda_n$ in Eq.~(\ref{eq:apRec}) to $\avLambda{n}$, and then only stage (ii) remains:
\begin{equation}
\avLambda{n+1} =
\left\langle
\tilde U_1^\dagger W_0^\dagger \tilde U_1
\,\,\avLambda{n}\,\,
\tilde U_1^\dagger W_0 \Lambda \tilde U_1
\right\rangle_{\tilde U_1}
\,.
\end{equation}

Since the average $\avLambda{n}$ is isotropic with respect to the group, $\tilde U_1$ and $\tilde U_1^\dagger$ around it can be dropped, and we obtain Eq.~(\ref{eq:LamRec}) (cf. the definition (\ref{eq:LamU}), (\ref{eq:ave})).

\section{Gate family with $m_1=m_2=1/5$}
\label{sec:appGateFamily}

We note a special role of the gate family with $m_1=m_2=1/5$. In this case the second and third eigenvalues of $M^0$ in (\ref{eq:Sp}) vanish, and the decay curve is a single exponential to a high accuracy. In other words, these gates are especially suited for partial randomized benchmarking. Furthermore, for these gates next-order corrections~(\ref{eq:2ndorder}) also vanish.
Such a gate can be implemented, for example, by turning on the qubit-qubit coupling $H_\textrm{cpl}=g[1-(1+\lambda)\sigma_x^1\sigma_x^2
-(1-\lambda)\sigma_y^1\sigma_y^2 - \sigma_z^1\sigma_z^2]$ for a finite period $t=h/(16g)$, provided that $\cos\lambda\pi=-1/5$:
\begin{equation}\label{eq:5fam}
W_\lambda = \exp \left( -\frac{i}{\hbar} H_\textrm{cpl}\,t \right) =
\begin{pmatrix}
\cos\frac{\lambda\pi}{4}&0&0&i\sin\frac{\lambda\pi}{4}\\
0&\frac{1-i}{2}&\frac{1+i}{2}&0\\
0&\frac{1+i}{2}&\frac{1-i}{2}&0\\
i\sin\frac{\lambda\pi}{4}&0&0&\cos\frac{\lambda\pi}{4}
\end{pmatrix}
\,.
\end{equation}
For general $\lambda$ the gates in Eq.~(\ref{eq:5fam}) describe a family with $m_1=m_2= \frac{1}{24} (5+\cos\lambda\pi)$ between $1/6$ and $1/4$.

\section*{Acknowledgments}

We thank A.~Shnirman for valuable discussions. This work was supported via the Basic research program of HSE.

\bibliographystyle{apsrev}
\bibliography{rb}

\end{document}